\documentclass[conference]{IEEEtran}
\IEEEoverridecommandlockouts
\usepackage{cite}
\usepackage{amsmath,amssymb,amsfonts}
\usepackage{multirow}
\usepackage{algorithmic}
\usepackage{graphicx}
\usepackage{subcaption}
\usepackage{textcomp}
\usepackage{xcolor}
\usepackage{hyperref}

\def\BibTeX{{\rm B\kern-.05em{\sc i\kern-.025em b}\kern-.08em
    T\kern-.1667em\lower.7ex\hbox{E}\kern-.125emX}}
        
\usepackage{fancyhdr}
\begin{document}

\title{How unitizing affects annotation of cohesion\\
\thanks{This paper has been partially supported by the French National Agency (ANR) in the frame of its Technological Research JCJC program (GRACE, project ANR-18-CE33-0003, funded under the Artificial Intelligence Plan).  \\
\newline
\textcopyright 2019 IEEE. Personal use of this material is permitted. Permission from IEEE must be obtained for all other uses, in any current or future media, including reprinting/republishing this material for advertising or promotional purposes, creating new collective works, for resale or redistribution to servers or lists, or reuse of any copyrighted component of this work in other works. \\
\newline
DOI: 10.1109/ACII.2019.8925527
}
}

\author{\IEEEauthorblockN{Eleonora Ceccaldi}
\IEEEauthorblockA{\textit{Casa Paganini - InfoMus, DIBRIS} \\
\textit{University of Genoa}\\
Genoa, Italy \\
eleonora.ceccaldi@edu.unige.it}
\and
\IEEEauthorblockN{Nale Lehmann-Willenbrock}
\IEEEauthorblockA{\textit{Dept. of Industrial/Organizational Psychology} \\
\textit{University of Hamburg}\\
Hamburg, Germany \\
nale.lehmann-willenbrock@uni-hamburg.de}
\and
\IEEEauthorblockN{Erica Volta}
\IEEEauthorblockA{\textit{Casa Paganini - InfoMus, DIBRIS} \\
\textit{University of Genoa}\\
Genoa, Italy \\
erica.volta@edu.unige.it}
\and
\IEEEauthorblockN{Mohamed Chetouani}
\IEEEauthorblockA{\textit{ISIR
%Institut des Syst\`{e}mes Intelligents et de Robotique
} \\
\textit{Sorbonne Universit\'{e}, CNRS UMR 7222}\\
Paris, France \\
mohamed.chetouani@sorbonne-universite.fr}
\and
\IEEEauthorblockN{Gualtiero Volpe}
\IEEEauthorblockA{\textit{Casa Paganini - InfoMus, DIBRIS} \\
\textit{University of Genoa}\\
Genoa, Italy \\
gualtiero.volpe@unige.it}
\and
\IEEEauthorblockN{Giovanna Varni}
\IEEEauthorblockA{\textit{LTCI} \\
\textit{T\'el\'ecom Paris, Institut polytechnique de Paris}\\
Paris, France \\
giovanna.varni@telecom-paristech.fr}
}

\maketitle
\thispagestyle{fancy}

\begin{abstract}
 This paper investigates how unitizing affects external observers' annotation of group cohesion. We compared unitizing techniques belonging to these categories: interval coding, continuous coding, and a technique inspired by a cognitive theory on event perception. We applied such techniques for sampling coding units from a set of recordings of social interactions rich in behaviors related to cohesion. Then, we compared the cohesion scores the observers assigned to each coding unit. Results show that the three techniques can lead to suitable ratings and that the technique inspired to cognitive theories leads to scores reflecting variability in cohesion better than the other ones.
\end{abstract}

\begin{IEEEkeywords}
Social interaction, cohesion, unitizing.
\end{IEEEkeywords}

\section{Introduction}
Social interactions elicit emotions that have the power of shaping and connecting groups \cite{van2017emotional}. Previously collected data-sets on emotions (e.g., \cite{correa2018amigos}) considered the social context as a key factor in affect and mood elicitation. Affective computing techniques were developed for group-level emotion recognition (e.g., \cite{gupta2018}). Moreover, organizational psychology has highlighted the relevance of group affect for explaining emergent team states, such as trust, conflict, and cohesion (e.g., \cite{barsade2015group, okhuysen2016}).
 To capture dynamic group phenomena, research often relies on behavioral coding schemes (see \cite{brauner2018} for an overview). These are systems for observing behavior, measuring their occurrence (e.g., number of smiles) or intensity (e.g., warmth) \cite{bakeman1997observing}. According to Reis and Judd \cite{reis2000handbook}, the most fundamental property of a coding scheme for observing social interaction is the technique adopted for sampling behavior, otherwise known as \textit{unitizing}. Unitizing divides an observation into discrete smaller coding units. The adopted unitizing technique may vary depending on the goal of the study, on technical constraints, or on the specific research question at hand \cite{lehmann-Willenbrock2018}.
 
 In social interaction, unitizing is both a crucial and complex task: it needs to take into account individual behaviors of single members of the group as well as group behavior and associated emergent states. In the perspective of endowing artificial agents with social intelligence, unitizing has relevant implications. We expect that agents able to segment interaction in meaningful units will be able to reach a comprehensive understanding of the interaction itself and to adequately plan their social behavior. Moreover, a proper unitizing will enable developing computational models that can catch the multiple facets of the social phenomena under investigation.   
 
 In this paper, we investigate how unitizing affects external observers' annotation of group cohesion. This emergent state was chosen because it is a multidimensional construct having five recognized dimensions (task, social, belongingness, group pride, and morale) \cite{salas2015}, involving both emotions \cite{magee2006emotional} and goals \cite{levine2018finding}. Moreover, cohesion was addressed by studies adopting different unitizing techniques and concerning both psychological and computational aspects. Whilst psychologists and sociologists investigated all five dimensions, at the present computer scientists started to address computational methods for automatic analysis of the task and social dimensions (e.g., see \cite{hung2010estimating, nanninga2017}). Our work further contributes to this research direction.
 Specifically, we applied three different unitizing techniques for sampling coding units from a dataset of audiovisual recordings and we compared the scores external observers assigned to each unit for annotating the task and social dimensions of cohesion. The first unitizing technique belongs to the family of \emph{interval coding} \cite{waller2018systematic}, the second one to the family of \emph{continuous coding} \cite{waller2018systematic}, and the third one is a technique inspired by the Event Segmentation Theory \cite{zacks2007event} in Cognitive Science, we conceived an instance thereof.
 
 \section{Background} 
\label{sec:background}
 When deciding upon the kind of unitizing to be adopted, Meinecke and colleagues \cite{meinecke2015social} suggest answering the following question: are behavioral codes assigned to a behavioral event or are codes assigned to a specific time interval?

 One viable option is \textit{interval coding} \cite{waller2018systematic}, standing for identifying a fixed-length interval of time for raters to note the occurrence of any of target behavior. \emph{Thin slices} \cite{ambady1992} is a well-know approach of this family and extensively used in social psychology and computational studies. According to it, fixed-length time windows of behavior from 2 seconds to 5 minutes are deemed to provide an efficient assessment of personality, affect, and interpersonal relations.
 Gatica Perez and colleagues used this approach to detect high interest level in meetings \cite{gatica2005detecting}. Hung and colleagues for the first time adopted thin slices for automated estimation of cohesion \cite{hung2010estimating}. Interval coding has many advantages \cite{bakeman1997observing, gatica2005detecting}: it is fast and it can be easily automatized so that it is less prone to the experimenter's subjectivity. Moreover, it does not require a prior knowledge of the content of an interaction. Nevertheless, it might lead to boundaries of the coding units being placed within actions, thus resulting in losing important information. For example, it might split an interaction before its ending (e.g., think of 2 people interrupted as they speak or of a smile following a seemingly harsh statement: it might overturn the meaning of it, but would the statement and the smile end up into different segments, such overturning will be lost). 
 
 Another option is \textit{continuous coding} \cite{waller2018systematic}, meaning that every single utterance, gesture, and so on is annotated. Also, it is possible to look and sample for specific behaviors, either at scheduled or random points, throughout a recording. Therefore, in this case, the unitizing step has the role of identifying the units of analysis. Continuous coding mostly requires a coding scheme itself. Among those for group interaction, \emph{ACT4Teams} -- a coding scheme for measuring problem-solving dynamics in groups, see \cite{kauffeld2018} -- breaks the observation into \textit{thought units}, i.e., the smallest meaningful segments of behavior that can be coded into the $43$ categories composing this coding scheme. ACT4Teams was used to annotate social and task cohesion from verbal content in \cite{nanninga2017}.
 Despite the convenience of being tailored for a specific objective or approach, continuous coding is time consuming and often requires annotators to be trained on the methodology. 
 
 According to cognitive sciences \cite{zacks2007event}, unitizing is an automatic component of human perceptual processing of the ongoing situation. In other words, humans make sense of reality by breaking it into smaller units. The Event Segmentation Theory (EST) \cite{zacks2007event} leverages the innate ability of human beings to parse an ongoing interaction into meaningful units. Research showed that even 10 months old infants can perceive action boundaries \cite{hespos2010infants}. Studies, e.g., \cite{newtson1976perceptual}, also demonstrated that the perceptual organization of ongoing behavior can be measured by asking participants to watch a video-clip depicting daily life activities and to press a key whenever they feel \emph{``a meaningful portion of action ends and another one begins"}. Zacks \cite{zacks2001event} defines an event as \emph{``a segment of time at a given location, that is conceived by an observer to have a beginning and an end''}. Through the process of \textit{``event structure perception"} \cite{zacks2001event} discrete units are perceived as segments of time having a beginning and an end. EST provides a computational and neurophysiological account of event structure perception. According to it, event segmentation relies on event models observers form of the ongoing situation. Event models are cognitive representations of the current state of affairs, based on perception and previous experience. Such models frame new incoming information and guide prediction of future developments. The perception of event boundaries is closely tied to prediction: a boundary is perceived whenever unpredictable changes occur, putting the currently active event model at stake. Behavioral data in \cite{zacks2009segmentation} demonstrated the association between changes and boundary perception. Boundaries were identified as a consequence of changes in 7 dimensions: (1) \emph{time}, (2) \emph{space}, (3) \emph{objects}, (4) \emph{characters}, (5) \emph{character interaction}, (6) \emph{causes}, (7) \emph{goals}. Moreover, effect is incremental: the more situational features change, the larger the probability that a viewer would identify an event boundary.
 
 Notwithstanding the increased understanding of pros and cons of the different unitizing techniques, a systematic comparison of them for analysis of social interaction is missing. Moreover, EST was never applied for this purpose. This paper addresses these issues using cohesion as a test-bed.

 \section{An EST-inspired methodology} 
\label{sec:EST}
 As shown by the EST, humans make sense of experience by dividing it into meaningful units that are based on cognitive models of the ongoing situation. Like an horizontal and a vertical line are effortlessly perceived as a ``T'', putting hot water into a cup, placing a tea-bag in it, and waiting for a few minutes are grouped into a ``making tea'' event. The updating of event models depends on changes in the situation. In the ``tea'' example, one might predict sugar being added to the tea. This ``fortune telling'' function of event models lasts until it is no longer possible to predict the future unfolding of events by relying on the same representation. A jumping-jack performed after the tea has been drunk might be difficult to integrate into the ``making tea'' event. This might indicate the need of reshaping the model, 
 %so that it can be updated with the new information, 
 or else it might imply that the ``making tea'' event has ended and a new event has begun. From this theory, we conceived a technique for unitizing interaction manually. It consists of the following steps:
 \begin{itemize}
    \item \textit{Step 1 -- Annotating changes}: annotating each change in the scene that can be related to one of the 7 categories. The first step would be parsing the material and keeping track of each change category that can be distinguished.
    \item \textit{Step 2 -- Placing boundaries}: detecting boundaries based on the changes in the scene. According to \cite{zacks2009segmentation}, changes in different categories correlate with boundary perception, with a spike in the probability of a boundary being detected after three changes. For this reason, we cut the scene after changes in three different categories are found. What is more, Levine and colleagues \cite{levine2018finding} illustrate how goals elicit boundary perception more than other lower-level changes. As a consequence, in our technique, changes belonging to the ``goal'' category value twice.
 \end{itemize}
 
 \begin{table*}[htb!]
\centering
\small
\caption{The left column reports the changes categories as described in \cite{zacks2009segmentation}. The column in the middle shows how these changes were operationalized in our EST-based technique. The right column provides an example of occurrence of each change.}
\begin{tabular}{|l|l|l|}
\hline
\bf{Changes in EST} & \bf{Changes in our EST-inspired technique} & \bf{Example in group interaction}\\
\hline
C1. Time & Timing and rhythm of the interaction & Group members start gesticulating fast\\
\hline
C2. Space & Motion direction & Group members all move their heads towards the speaker \\
\hline
C3. Objects & Interaction with objects & Participants dismiss a tool they were using\\
\hline
C4. Characters & Character location & One group member leaves\\
\hline
C5. Character interaction & Interaction patterns & Group members start mocking each-other\\
\hline
C6. Causes & Causes and appraisal & Something happens as a consequence of a new state of affairs\\
\hline
C7. Goals & Goals fulfilled, dismissed, or replaced & Group members stop paying attention to the speaker\\
\hline
\end{tabular}
\label{ref:tableChanges}
\end{table*}
 
 Our technique derives boundaries from changes in $7$ situational dimensions, inspired by the EST but fine-tuned for unitizing group interaction. The most remarkable difference regards the ``time'' category. In our technique, time is conceptualized as the timing, or rhythm, of the interaction, instead of changes in temporal reference of the scene. Table \ref{ref:tableChanges} illustrates how we operationalized each change category.
 As a result, we expect our unitizing technique to provide segments that are perceived as more natural and understandable by annotators and coders. Moreover, we hypothesize that the unitizing technique has an effect on the quality of the annotation.
 
 \section{Experiment}
\label{sec:expe}
 We conducted an online perceptual experiment to compare three unitizing techniques with respect to how they affect:
 \begin{itemize}
    \item The average agreement on the raters' scores;
    \item The extent to which these scores reflect the intrinsic variability of the data-set; 
    \item The loss of information with respect to the scores an expert rater gave to the whole (non-unitized) interactions. That is, a rater scoring a coding unit does it on the basis of information which is limited with respect to a rater who scores the whole interaction \cite{guetzkow1950unitizing}.
 \end{itemize}
 To this aim, a pool of external observers rated task and social dimensions of cohesion on coding units obtained by applying the three different unitizing techniques.

\subsection{Stimuli}
 The stimuli used in this study are audio-video recordings from the Panoptic data-set \cite{Joo_2017_TPAMI}. This is a multimodal publicly available data-set with the following features: natural interactions having rich and subtle non-verbal cues, small groups of up to 8 people, and a large number of camera views (up to 521). The data-set includes recordings of people playing several social games in small groups. In this study, we focused on the \emph{Ultimatum} game. This game is often used in experimental economics and psychology to study conflict and cooperation. Rules are very simple: one or more players, the \emph{proposers}, are given a sum of money to be split with another player(s), the \emph{responders}. Proposers and responders have a limited amount of time to discuss on how to split the money starting from a proposal done by the proposers. At the end of the established time, if the players agreed on the split, they gain the money, otherwise they loose it. 
 The Panoptic data-set includes 5 Ultimatum's sessions involving each one from 3 to 8 players. A visual inspection of the sessions was performed by an expert psychologist. This resulted in the selection of 12 interactions displaying a variety of behaviors related to the social and task dimensions of cohesion and spanning a broad range of cohesion intensity. Selected interactions involved 3 to 7 players. Interactions were 46 to 57 seconds long (M=53s, SD=3.3s). We discarded interactions involving 8 players because due to the large amount of persons simultaneously acting in the scene, occlusions often occurred making the job of the raters difficult.

\subsection{Unitizing}
\begin{figure}[ht]
 \centering
    \includegraphics[width=0.33\textwidth]{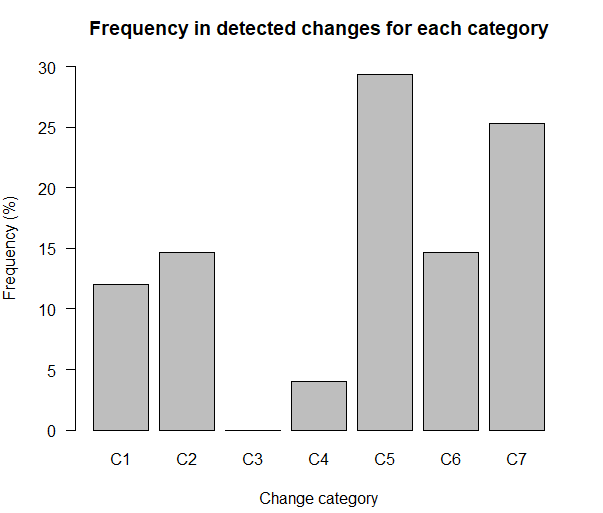}
    \caption\footnotesize{Frequency of the 7 EST change categories in the data-set.}
    \label{bar_chart}
 \end{figure}

\begin{figure*}[ht]
  \centering
  \begin{subfigure}[b]{0.3\linewidth}
    \includegraphics[width=\linewidth]{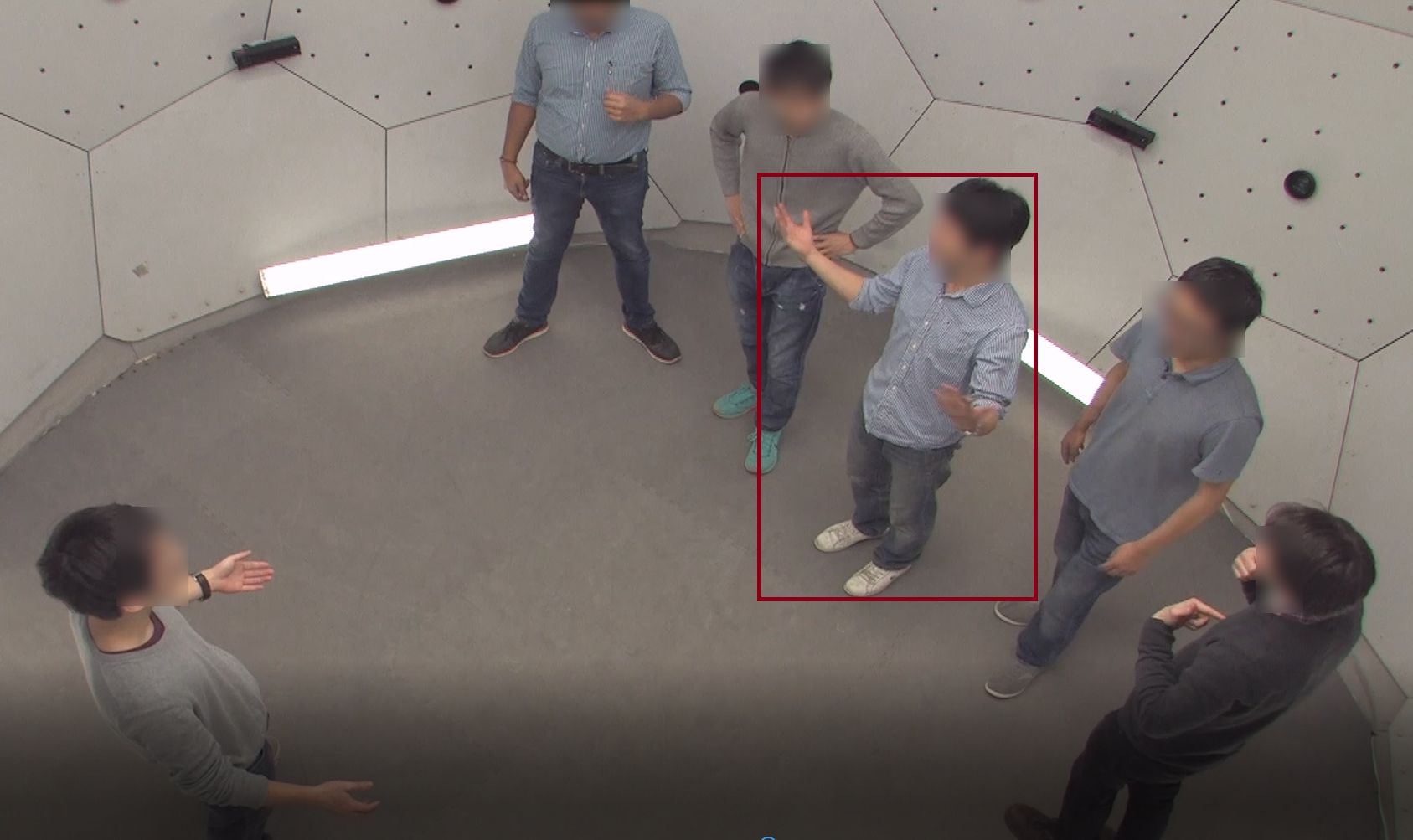}
     \caption{}
  \end{subfigure}
  \begin{subfigure}[b]{0.3\linewidth}
    \includegraphics[width=\linewidth]{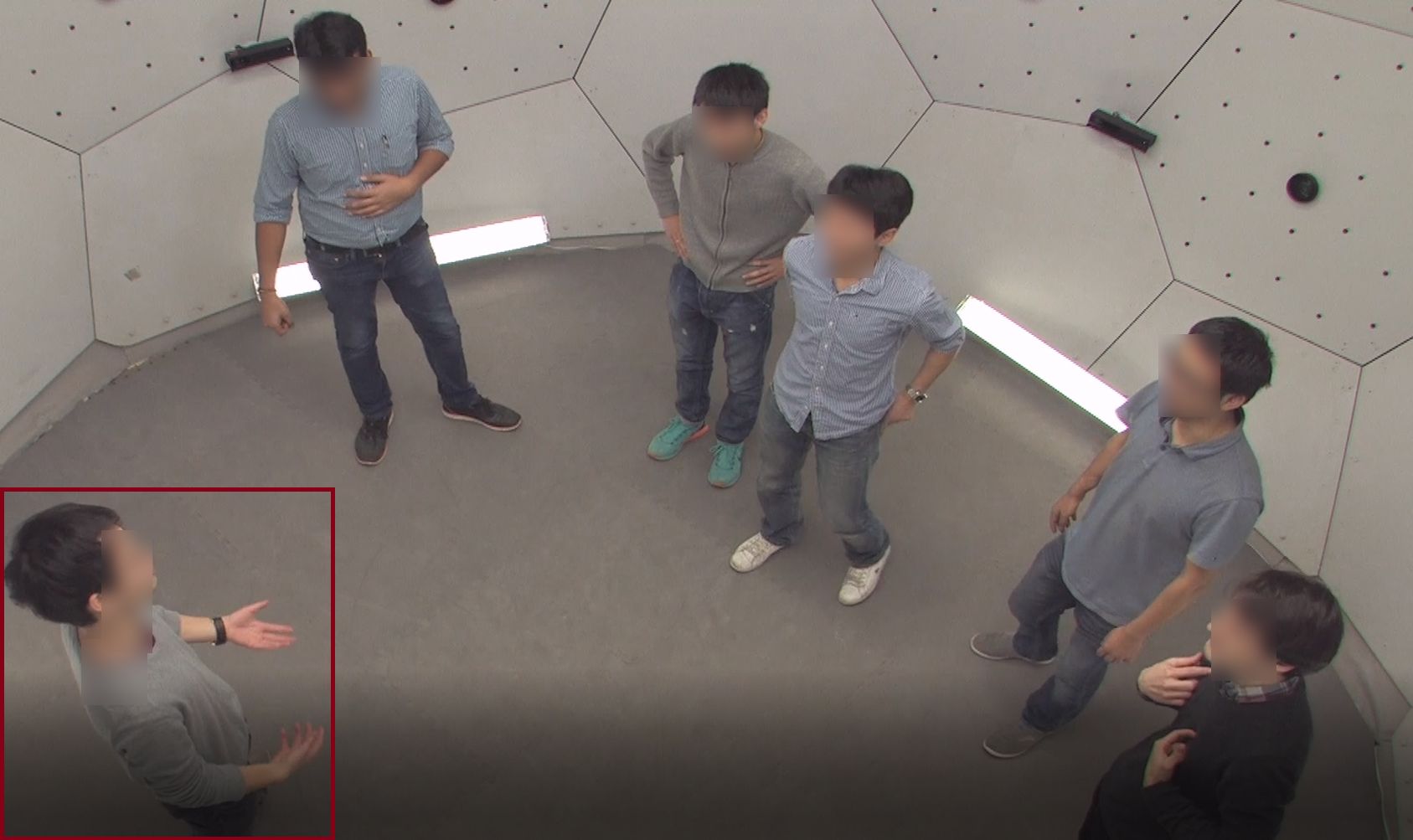}
    \caption{}
  \end{subfigure}
  \begin{subfigure}[b]{0.3\linewidth}
    \includegraphics[width=\linewidth]{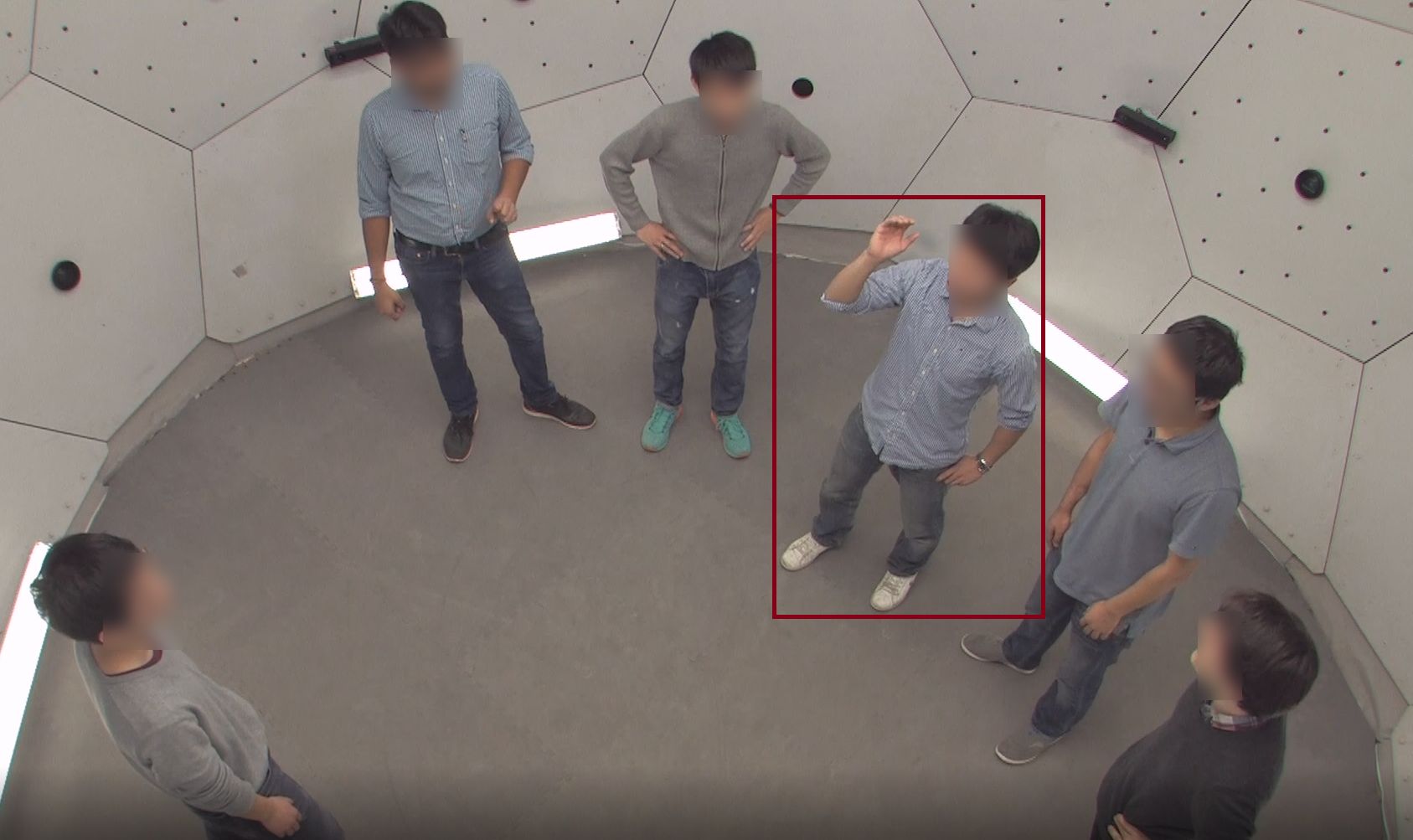}
    \caption{}
  \end{subfigure}
    \begin{subfigure}[b]{0.3\linewidth}
    \includegraphics[width=\linewidth]{event3_censored.jpg}
    \caption{}
  \end{subfigure}
  \begin{subfigure}[b]{0.3\linewidth}
    \includegraphics[width=\linewidth]{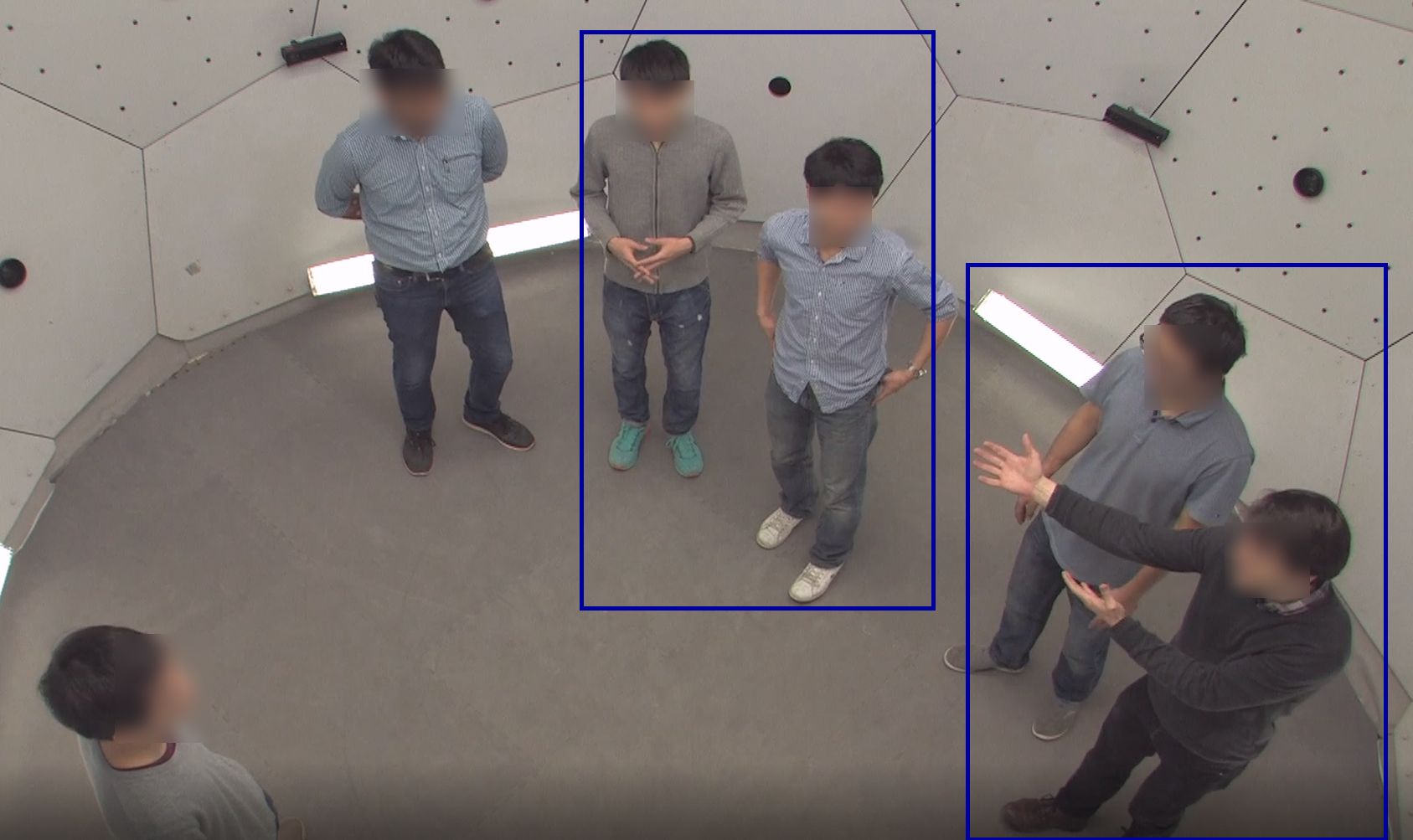}
    \caption{}
  \end{subfigure}
    \begin{subfigure}[b]{0.3\linewidth}
    \includegraphics[width=\linewidth]{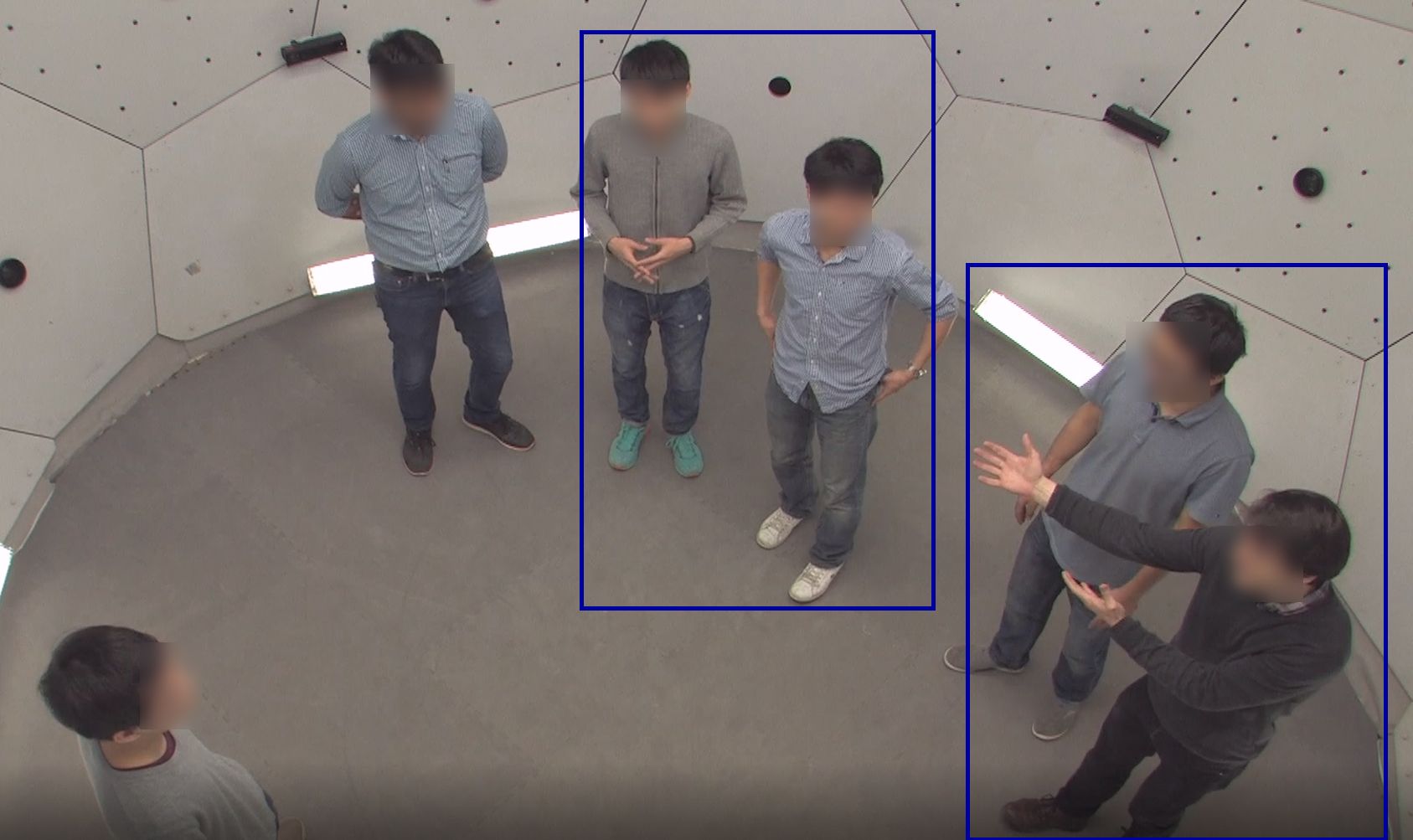}
    \caption{}
  \end{subfigure}
  \caption\footnotesize{Example of EST unitizing. In each panel, a bounding box identifies the changes in the scene. In the upper row, only the speaker changes from frame (a) to (b) and from frame (b) to (c). The changes all fall into the same category, i.e., character interaction (red), therefore the frames all belong to the same event. In the lower row, three different categories of changes are observed: from frame (d) to frame (e) the speaker changes (red), the motion direction changes (blue), and the goals of the players change as one of them joking distracts other players from the game (pink). Hence a boundary is placed between frame (d) and (e). Again, only one type of change (movement) is observed between frame (e) and (f), belonging to the same event.} 
\label{fig:segmentation}
\end{figure*}

 To address continuous coding we selected the unitizing technique specified in the ACT4Teams coding scheme (\emph{ACT}). A new unit is created whenever the speaker changes, whenever a speaker utters several statements expressing a thought, whenever the main argument changes, or whenever the speaker talks for longer than 20 seconds. Videos were manually parsed to generate coding units according to this technique.
 Concerning EST, a trained psychologist watched each interaction and unitized it according to the technique described in Section \ref{sec:EST} (\emph{EST}). Fig. \ref{fig:segmentation} displays how changes were used to unitize the interaction. As previously described, a new coding unit is created whenever 3 changes are detected in the interaction. In Fig. \ref{fig:segmentation}, Panels (a), (b) and (c) belong to the same event: the speaker changes from the first panel to the second and again in the third. Such changes all belong to the \textit{character interaction} category, whereas no change is detected in the other categories. On the contrary, three changes occur between Panels (c) and (d), thus originating a boundary: all participants have changed their position, turning their head towards the player in the right corner. Also, the frame portrays the players while one of them is telling a joke, making everyone laugh, thus resulting in a change in the character interaction. Furthermore, the main goal in the scene has changed, as players have stopped being focused on the game. Panels (d), (e), and (f) belong again to the same event. The bar chart in Fig. \ref{bar_chart} reports the percentage of changes in our data-set for each category (no changes were detected for the \textit{object interaction} category).
 
 Concerning interval coding, we adopted three different sizes for the fixed-window: 8s (\emph{AUT8}), 15s (\emph{AUT15}), and 21s (\emph{AUT21}), respectively. These values were chosen by taking into account previous work on analysis of social interaction in small groups (15s as in \cite{gatica2005detecting}), and the average duration of the coding units obtained by applying ACT (8s) and EST (21s).  
 Table \ref{ref:segments/methodology} summarizes the results of this procedure. 

\begin{table}[ht!]
    \footnotesize
    \centering
    \caption{Summary of the output of the unitizing procedure.}
    \begin{tabular}{|c|c|c|} %{|p{2.5cm}|p{2.5cm}| p{2.5cm}|}
        \hline \centering
        \bf{Unitizing} & \bf{No. of coding units} & \bf{Avg duration and SD [s]}\\
        \hline \centering 
        EST & $30$ & M = $21$ (SD = $10$)\\
        \hline \centering
        ACT & $61$ & M = $8$ (SD = $5$)\\
        \hline \centering
        AUT8 &  $73$ & M = $8$ (SD = $0$)\\
        \hline \centering
        AUT15 &  $47$ & M = $15$ (SD = $0$)\\
        \hline \centering
        AUT21 &  $33$ & M = $21$ (SD = $0$)\\
        \hline
    \end{tabular}
    \label{ref:segments/methodology}
 \end{table}

\subsection{Questionnaire}
\label{quest}
 The conceptual model of cohesion by \cite{brawley1987assessing}, originally applied to sport contexts, is at the basis of most questionnaires used in group cohesion research (e.g., \cite{estabrooks2000physical, hung2010estimating, chang2014high, chin1999perceived, warkentin1997virtual}). These questionnaires enable the assessment of cohesion over its dimensions both in first and in third person (i.e., as self-perception and from the point of view of an external rater). We pooled together items from \cite{hung2010estimating, warkentin1997virtual, brawley1987assessing} to create a questionnaire containing 10 items organized in two subscales, concerning the task and social dimensions of cohesion. The Likert items of the adopted questionnaires consisted of 7 \cite{hung2010estimating, warkentin1997virtual} and 9 \cite{brawley1987assessing} points, respectively. In this study, we adopted 5-points Likert items -- from 1 (\emph{Not at all}) to 5 (\emph{Yes, definitely}) -- as previous studies argued how a 5-points scale can make reading all answers easier for the responders \cite{dawes2008data}. Table \ref{ref:table} reports the scale used and provides the source for each item. Item from \cite{brawley1987assessing} was reported in third person and it was inverted. 

\begin{table*}[ht!]
\centering
\footnotesize
\caption{The questionnaire on cohesion used in this study.}
\begin{tabular}{|p{9.2cm}|p{8.0cm}|}
\hline
\bf{Task dimension} & \bf{Social dimension}\\
\hline
Do you feel that group members share the same purpose, goal, intentions? \cite{hung2010estimating} & Were group members open and frank in expressing ideas/feelings? \cite{warkentin1997virtual} \\
\hline
Do group members give each other a lot of feedback? \cite{hung2010estimating} & How engaged in the discussion do group members seem? \cite{hung2010estimating} \\
\hline
Do group members seem to have sufficient time to make their contribution? \cite{hung2010estimating} & Do group members appear to be in tune/in sync with each other? \cite{hung2010estimating} \\
\hline
Do group members have conflicting aspirations for the team's performance? \cite{brawley1987assessing} & Do group members listen attentively to each other? \cite{hung2010estimating} \\
\hline
Do group members respect individual differences and contributions? \cite{warkentin1997virtual}  & Does the group seem to share responsibility for the task? \cite{hung2010estimating} \\
\hline
\end{tabular}
\label{ref:table}
\end{table*}

\subsection{Participants}
 Ninety-nine persons (37 males, 60 females, 2 preferred not to specify their gender) voluntarily participated in the study. They were mainly recruited via email advertisements at several universities and research centers.

\subsection{Procedure}
 Ratings were collected via a web application in an anonymous form. First, raters were welcome, and given the instructions for their task. Then, they were asked to enter information about age and gender. No information about the Internet connection (e.g., IP address) was tracked nor collected. Participants were informed on data protection policies. Finally, they started to watch the coding units to be rated according to the items in Table \ref{ref:table}. Coding units were randomly administered. Next to the video showing the coding unit, a screenshot displayed the group of people to focus on. Raters could leave the experiment when they wished. The total number of gathered ratings was 1771, we discarded 23 ratings due to wrong answers to the honey pot questions appearing in the questionnaire each 10 coding units we asked to annotate. Such questions were used to detected whether the rater was still paying attention to the task. The average number of ratings for each rater was 15. 
 Some raters, some days after taking part in the experiment, contacted the experimenter to provide their feedback about the experience. In particular, most of them reported having had difficulties to fully understand the following two items of the questionnaire: \emph{Do group members respect individual differences and contributions?}, and \emph{Does the group seem to share responsibility for the task?}. For this reason, in this study, we decided to remove these items for running the analysis.
 
 \section{Analysis}
\label{sec:analysis}
\subsection{Effect of unitizing on raters' agreement}
 Inter-raters reliability was assessed\footnote{ The significance threshold for all the tests in this study was set at .05} by means of a one-way average-measures (ICC) to evaluate whether raters agreed in their ratings of cohesion across unitizing techniques (AUT8, AUT15, AUT21, ACT, EST). Table \ref{ref:ICC} shows the obtained values of ICC and the respective 95\%-confidence intervals. All the resulting ICCs were in the \emph{excellent} range \cite{koo2016}, indicating that raters had a high degree of agreement and that both the task and the social dimensions of cohesion were similarly rated across raters independently by the unitizing technique.
 
\begin{table}[htbp]
\centering
\caption{ICC values and Confidence Intervals for the task and the social dimension of cohesion}
\begin{tabular}{|l|l|l|}
\hline
\multicolumn{1}{|c|}{\multirow{2}{*}{\bf{Unitizing}}} & \multicolumn{2}{c|}{\bf{ICC and 95\%-CI}}                                \\ \cline{2-3} 
\multicolumn{1}{|c|}{}                                      & \multicolumn{1}{c|}{\bf{Task}} & \multicolumn{1}{c|}{\bf{Social}} \\ \hline
\bf{AUT8}     & 0.96 (95\% CI [0.94, 0.97])   &    0.94 (95\% CI [0.92, 0.96])            \\ \hline
\bf{AUT15}    & 0.95 (95\% CI [0.93, 0.97])   &   0.97 (95\% CI [0.95, 0.98])               \\ \hline
\bf{AUT21}    & 0.98 (95\% CI [0.97, 0.99])   &   0.96 (95\% CI [0.94, 0.98])                \\ \hline
\bf{ACT}      & 0.96 (95\% CI [0.95, 0.97])   &    0.97 (95\% CI [0.95, 0.98])                 \\ \hline
\bf{EST}      & 0.96 (95\% CI [0.94, 0.98])   &    0.97 (95\% CI [0.95, 0.98])                  \\ \hline
\end{tabular}
\label{ref:ICC}
\end{table}

 The three unitizing techniques (including the three instances of interval coding) can be therefore considered a viable option to achieve suitable ratings for use in hypothesis tests on  social and task dimensions of cohesion.
 For further analysis, the average of the raters’ scores is assigned to each coding unit.

\subsection{Effect of unitizing on cohesion scores}
\label{sec:analysis_scores}
 To assess the extent to which cohesion scores reflect the variability of the data-set, we analyzed the variance of the scores for each unitizing technique. For both dimensions of cohesion, we first checked for differences between the three instances of interval coding techniques (AUT8, AUT15, and AUT21). Then we compared the interval coding technique which better reflects variability with ACT and EST.

\subsubsection{Comparison between interval coding techniques}
 \emph{\\Task}: A Brown-Forsythe test detected a significant effect of unitizing technique  (F(2,85.06)=5.43, p=.006). Post hoc comparisons detected a significant difference for AUT21 (SD=2.49) and AUT08 (SD=1.99), p=.015, and for AUT21 and AUT15 (SD=1.51), p=.005. No significant difference was found between AUT8 and AUT15 (p=.52). A Bonferroni correction was applied to account for multiple comparisons. \\
 \emph{Social}: A Brown-Forsythe test was conducted. A significant effect of unitizing technique was found (F(2,113.67)=27.20, p$<$.001). Post hoc comparisons (Bonferroni correction applied) detected a significant difference for AUT21 (SD=1.61) and AUT08 (SD=1.5), p$<$.001, and for AUT21 and AUT15 (SD=1.86), p$<$.001, and for AUT8 and AUT15 (p$<$.001). \\
 Results show that AUT21 reflects variability better than AUT08 and AUT15, having the highest SD for the task dimension and being comparable with AUT15 for the social one. We therefore retained AUT21 for subsequent analysis.
 
\subsubsection{Comparison between AUT21, ACT, and EST}
 \emph{\\Task}: A Brown-Forsythe test was run to compare variances of ACT, AUT21, and EST. A significant effect of unitizing technique was found (F(2,70.38)=12.48, p$<$.001). Post hoc comparisons (Bonferroni correction applied) detected a significant difference for ACT (SD=1.28) and AUT21 (SD=2.49), p$<$.001, and for ACT and EST (SD=1.72), p=.003. No significant difference was found between AUT21 and EST (p=.07).\\
 \emph{Social}: A Brown-Forsythe test was run and a significant effect was found (F(2,104.91)=22.81, p$<$.001). Post hoc comparisons (Bonferroni correction applied) detected a significant difference for ACT (SD=2.15) and AUT21 (SD=1.61), p$<$.001, and for AUT21 and EST (SD=1.89), p$<$.001. No significant difference was found between ACT and EST (p=.055). \\
 Results show that EST and AUT21 reflect variability in the same way and better than ACT for the task dimension. Concerning the social one, there is no significant difference between EST and ACT, and both outperform AUT21. EST thus best reflects variability in both dimensions of cohesion.

\subsection{Effect of unitizing on loss of information}
 To assess the effect of unitizing technique on loss of information, we compared the scores obtained with each technique with the scores provided by an expert rater who watched the whole non-unitized interaction. For each interaction, Mean Square Error (MSE) was computed between these scores. According to the findings in Section \ref{sec:analysis_scores}, we carried out this analysis on ACT, AUT21, and EST.

\subsubsection{Task} 
 Due to a deviation from a normal distribution of MSE for AUT21 (Shapiro-Wilk test, W=.80, p=.02), a Kruskal-Wallis test was conducted to examine the differences on MSE according to the unitizing technique. No significant difference (${\chi}^2$=2.85, p=.24, df=2) was found. 
 
\subsubsection{Social} 
 MSE did not deviate from a normal distribution (Shapiro-Wilk test). A Bartlett test of homogeneity of variances indicated that the assumption of homoscedasticity had been violated  (p=.004). A Welch's ANOVA was therefore conducted. A significant effect of unitizing technique was found (F(2.0,14.5)=4.76, p=.03). Post hoc comparisons using the Games-Howell test indicated that MSE for ACT (M=25.41, SD=19.37) was significantly different than AUT21 (M=6.92, SD=5.45), p=.037. EST (M=13.80, SD=13.12) did not significantly differ from ACT and AUT21. Results show that for the social dimension of cohesion, ACT deviates most from the scores given to the non-unitized interaction. 
 
\subsection{Effect of unitizing on ranking of coding units}
  We decided to investigate in more depth the performances of the unitizing techniques by proceeding coding unit by coding unit. Concretely, we ranked the coding units in order of increasing standard deviation of the raters' scores. For each unitizing technique, we then computed curve $C$ representing the number of coding units falling below the $n$-th percentile of standard deviation for $n$ ranging from 5 to 100, step=5. Finally, we compared such a distribution with the ideal distribution where all the coding units generated by a unitizing technique occupy the first positions in the ranking. For performing the comparison, we computed the ratio between the area under curve $C$ and the area under the curve representing the ideal distribution. We obtained the following ratios: for the task dimension, 0.57 for EST and ACT, 0.74 for AUT21; for the social dimension, 0.60 for EST and AUT21, 0.65 for ACT.

\section{Discussion}
 Concerning agreement, analysis shows that all the unitizing techniques represent a viable option.
 Regarding cohesion scores, EST outranked the other techniques in reflecting variability of cohesion in the units for both dimensions. Whereas the task dimension is better assessed when longer units are observed (EST and AUT21 outperformed ACT), our results do not support the same idea for the social dimension, as shorter units (ACT) provided greater variability than longer ones. We think this can be ascribed to raters needing more time to figure out task dynamics than the social one from the players' behaviors. Indeed, raters did not know the rules of the Ultimatum game. For this reason, short coding units could appear not enough ``readable'' for raters in terms of assessing whether players share goals, intentions, and make contribution. Moreover, following \cite{rohlfing2019multimodal}, observing an instance of task-related behavior (e.g., turn-taking), requires at least two individual contributions lasting averagely 2s each, whereas emotion recognition can occur in a shorter time (300 ms can be enough) \cite{dhall2015video}. This confirms that whilst short units were not suitable for the task dimension (especially ACT), they could instead be leveraged for the social one. As a consequence, when trying to evaluate both dimensions of cohesion, a flexible (in terms of time-window) technique is expected to lead to better evaluations. In this study, EST's better performance in reflecting variability could be ascribed to the higher variability in units duration (M=21, SD=10), in line with the idea illustrated in \cite{zacks2009segmentation} of event perception as a flexible process, that can be fine or coarse grained according to the  scope and goals of the perceiver. 
 Concerning the ranking of the coding units, the first 10\% of entries (i.e., the first 12 entries) reflects the results discussed above. For the task dimension, 6 units belong to the AUT21 category, 5 belong to EST, and 1 to ACT. For the social dimension, 6 units belong to the AUT21 category, 3 belong to EST, and 3 to ACT.
 Interestingly, EST units are also diverse in the changes categories they contain with respect to the dimensions of cohesion. The same EST unit had the higher ranking (i.e., lower standard deviation) for both dimensions. This unit contained changes both in character goals and in their interaction. What is more, the next EST entries are not the same entries for the task (4 entries) and social (2 entries) dimensions. With all 4 task entries containing goal changes but only one also containing character interaction change, and both social entries containing changes in character interaction but not in character goals. This diversity in changes distribution aligns with the definition of task and social cohesion provided by \cite{carron2000cohesion}, with the former indicating group goals and objectives and the latter indicating group members concerns towards relationships within the group.
 Our results suggest that basing unitizing on the course of the interaction over time (i.e., on changes in the interaction), rather than on time only (AUT techniques) or on behaviors (ACT) can help tackle both the task and social dimensions of group cohesion. An automated EST-based unitizing should focus on such changes primarily. 

\section{Conclusion}
 In this study, we investigated how unitizing affects external observers' assessment of group cohesion. We compared unitizing techniques belonging to three different categories: interval coding, continuous coding, and a technique inspired by a cognitive theory \cite{zacks2007event} on event perception. 
 %We ran an experiment to evaluate the performances of such techniques on social interactions recordings rich in behaviours related to the task and social dimensions of cohesion. 
 Results show that all the techniques can lead to suitable ratings, and that the EST technique leads to scores which reflect more variability in cohesion in different interactions. %Although our study focused on cohesion, in the future the feasibility of the EST-based technique will be assessed in other aspects of group interaction. 
 Moreover, this work provides hints to implement an EST-based automated unitizing. As \cite{waller2018systematic} points out, automatizing group coding (or at least some parts of it) would lead to more efficient and accurate ratings.  
 
 This is the first study using EST for unitizing social interaction. Nevertheless, it only focused on a single emergent state, i.e., cohesion. In the future, we aim at assessing whether such unitizing technique is suitable for the annotation of other emergent states such as conflict, trust, and so on.

\section*{Acknowledgments}
We thank Hanbyul Joo for his precious help with the Panoptic dataset and Roberto Sagoleo for the useful discussions about the annotation procedure. 
%The preferred spelling of the word ``acknowledgment'' in America is without 
%an ``e'' after the ``g''. Avoid the stilted expression ``one of us (R. B. 
%G.) thanks $\ldots$''. Instead, try ``R. B. G. thanks$\ldots$''. Put sponsor 
%acknowledgments in theunnumbered footnote on the first page.
\bibliographystyle{IEEEtran}
\bibliography{main}

\begin{thebibliography}{10}
\providecommand{\url}[1]{#1}
\csname url@samestyle\endcsname
\providecommand{\newblock}{\relax}
\providecommand{\bibinfo}[2]{#2}
\providecommand{\BIBentrySTDinterwordspacing}{\spaceskip=0pt\relax}
\providecommand{\BIBentryALTinterwordstretchfactor}{4}
\providecommand{\BIBentryALTinterwordspacing}{\spaceskip=\fontdimen2\font plus
\BIBentryALTinterwordstretchfactor\fontdimen3\font minus
  \fontdimen4\font\relax}
\providecommand{\BIBforeignlanguage}[2]{{%
\expandafter\ifx\csname l@#1\endcsname\relax
\typeout{** WARNING: IEEEtran.bst: No hyphenation pattern has been}%
\typeout{** loaded for the language `#1'. Using the pattern for}%
\typeout{** the default language instead.}%
\else
\language=\csname l@#1\endcsname
\fi
#2}}
\providecommand{\BIBdecl}{\relax}
\BIBdecl

\bibitem{van2017emotional}
G.~A. van Kleef, M.~W. Heerdink, and A.~C. Homan, ``Emotional influence in
  groups: the dynamic nexus of affect, cognition, and behavior,'' \emph{Current
  opinion in psychology}, vol.~17, pp. 156--161, 2017.

\bibitem{correa2018amigos}
J.~A.~M. Correa, M.~K. Abadi, N.~Sebe, and I.~Patras, ``Amigos: A dataset for
  affect, personality and mood research on individuals and groups,'' \emph{IEEE
  Transactions on Affective Computing}, 2018.

\bibitem{gupta2018}
A.~Gupta, D.~Agrawal, H.~Chauhan, J.~Dolz, and M.~Pedersoli, ``An attention
  model for group-level emotion recognition,'' in \emph{Proceedings of the 20th
  ACM International Conference on Multimodal Interaction}, ser. ICMI '18.\hskip
  1em plus 0.5em minus 0.4em\relax New York, NY, USA: ACM, 2018, pp. 611--615.

\bibitem{barsade2015group}
S.~G. Barsade and A.~P. Knight, ``Group affect,'' \emph{Annual Review of
  Organizational Psychology and Organizational Behavior}, vol.~2, no.~1, pp.
  21--46, 2015.

\bibitem{okhuysen2016}
M.~J. Waller, G.~A. Okhuysen, and M.~Saghafian, ``Conceptualizing emergent
  states: A strategy to advance the study of group dynamics,'' \emph{The
  Academy Of Management Annals}, vol.~10, no.~1, pp. 561--598, 2016.

\bibitem{brauner2018}
E.~Brauner, M.~Boos, and M.~E. Kolbe, \emph{The Cambridge Handbook of group
  interaction analysis}.\hskip 1em plus 0.5em minus 0.4em\relax Cambridge
  University Press., 2018.

\bibitem{bakeman1997observing}
R.~Bakeman and J.~M. Gottman, \emph{Observing interaction: An introduction to
  sequential analysis}.\hskip 1em plus 0.5em minus 0.4em\relax Cambridge
  university press, 1997.

\bibitem{reis2000handbook}
H.~T. Reis and C.~M. Judd, \emph{Handbook of research methods in social and
  personality psychology}.\hskip 1em plus 0.5em minus 0.4em\relax Cambridge
  University Press, 2000.

\bibitem{lehmann-Willenbrock2018}
N.~Lehmann-Willenbrock and J.~A. Allen, ``Modeling temporal interaction
  dynamics in organizational settings,'' \emph{Journal of Business and
  Psychology}, vol.~33, pp. 325--344, 2018.

\bibitem{salas2015}
E.~Salas, R.~Grossman, A.~M. Hughes, and C.~W. Coultas, ``Measuring team
  cohesion: Observations from the science,'' \emph{Human factors}, vol.~57,
  no.~3, pp. 365--374, 2015.

\bibitem{magee2006emotional}
J.~C. Magee and L.~Z. Tiedens, ``Emotional ties that bind: The roles of valence
  and consistency of group emotion in inferences of cohesiveness and common
  fate,'' \emph{Personality and Social Psychology Bulletin}, vol.~32, pp.
  1703--1715, 2006.

\bibitem{levine2018finding}
D.~Levine, D.~Buchsbaum, K.~Hirsh-Pasek, and R.~Golinkoff, ``Finding events in
  a continuous world: A developmental account,'' \emph{Developmental
  psychobiology}, vol.~61, no.~3, pp. 376--389, 2018.

\bibitem{hung2010estimating}
H.~Hung and D.~Gatica-Perez, ``Estimating cohesion in small groups using
  audio-visual nonverbal behavior,'' \emph{IEEE Transactions on Multimedia},
  vol.~12, no.~6, pp. 563--575, 2010.

\bibitem{nanninga2017}
M.~C. Nanninga, Y.~Zhang, N.~Lehmann-Willenbrock, Z.~Szl\'{a}vik, and H.~Hung,
  ``Estimating verbal expressions of task and social cohesion in meetings by
  quantifying paralinguistic mimicry,'' in \emph{Proceedings of the 19th ACM
  International Conference on Multimodal Interaction}, ser. ICMI '17.\hskip 1em
  plus 0.5em minus 0.4em\relax New York, NY, USA: ACM, 2017, pp. 206--215.

\bibitem{waller2018systematic}
M.~J. Waller and S.~A. Kaplan, ``Systematic behavioral observation for emergent
  team phenomena: Key considerations for quantitative video-based approaches,''
  \emph{Organizational Research Methods}, vol.~21, no.~2, pp. 500--515, 2018.

\bibitem{zacks2007event}
J.~M. Zacks and K.~M. Swallow, ``Event segmentation,'' \emph{Current directions
  in psychological science}, vol.~16, no.~2, pp. 80--84, 2007.

\bibitem{meinecke2015social}
A.~L. Meinecke and N.~Lehmann-Willenbrock, \emph{Social Dynamics at Work:
  Meetings as a gateway}, ser. Cambridge Handbooks in Psychology.\hskip 1em
  plus 0.5em minus 0.4em\relax Cambridge University Press, 2015, pp.
  325–--356.

\bibitem{ambady1992}
N.~Ambady and R.~Rosenthal, ``Thin slices of expressive behavior as predictors
  of interpersonal consequences: A meta-analysis.'' \emph{Psychological
  bulletin}, vol. 111, no.~2, p. 256, 1992.

\bibitem{gatica2005detecting}
D.~Gatica-Perez, L.~McCowan, D.~Zhang, and S.~Bengio, ``Detecting group
  interest-level in meetings,'' in \emph{Proceedings IEEE International
  Conference on Acoustics, Speech, and Signal Processing, 2005.}, vol.~1.\hskip
  1em plus 0.5em minus 0.4em\relax IEEE, 2005, pp. I/489--I/492.

\bibitem{kauffeld2018}
S.~Kauffeld, N.~Lehmann-Willenbrock, and A.~L. Meinecke, \emph{The Advanced
  Interaction Analysis for Teams (act4teams) Coding Scheme}, ser. Cambridge
  Handbooks in Psychology.\hskip 1em plus 0.5em minus 0.4em\relax Cambridge
  University Press, 2018, pp. 422--–431.

\bibitem{hespos2010infants}
S.~J. Hespos, S.~R. Grossman, and M.~M. Saylor, ``Infants’ ability to parse
  continuous actions: Further evidence,'' \emph{Neural Networks}, vol.~23, no.
  8-9, pp. 1026--1032, 2010.

\bibitem{newtson1976perceptual}
D.~Newtson and G.~Engquist, ``The perceptual organization of ongoing
  behavior,'' \emph{Journal of Experimental Social Psychology}, vol.~12, no.~5,
  pp. 436--450, 1976.

\bibitem{zacks2001event}
J.~M. Zacks and B.~Tversky, ``Event structure in perception and conception.''
  \emph{Psychological bulletin}, vol. 127, no.~1, pp. 3--21, 2001.

\bibitem{zacks2009segmentation}
J.~M. Zacks, N.~K. Speer, and J.~R. Reynolds, ``Segmentation in reading and
  film comprehension.'' \emph{Journal of Experimental Psychology: General},
  vol. 138, no.~2, pp. 307--327, 2009.

\bibitem{guetzkow1950unitizing}
H.~Guetzkow, ``Unitizing and categorizing problems in coding qualitative
  data,'' \emph{Journal of Clinical Psychology}, vol.~6, no.~1, pp. 47--58,
  1950.

\bibitem{Joo_2017_TPAMI}
H.~Joo, T.~Simon, X.~Li, H.~Liu, L.~Tan, L.~Gui, S.~Banerjee, T.~S. Godisart,
  B.~Nabbe, I.~Matthews, T.~Kanade, S.~Nobuhara, and Y.~Sheikh, ``Panoptic
  studio: A massively multiview system for social interaction capture,''
  \emph{IEEE Transactions on Pattern Analysis and Machine Intelligence},
  vol.~41, no.~1, pp. 190--204, 2017.

\bibitem{brawley1987assessing}
L.~R. Brawley, A.~V. Carron, and W.~N. Widmeyer, ``Assessing the cohesion of
  teams: Validity of the group environment questionnaire,'' \emph{Journal of
  Sport Psychology}, vol.~9, no.~3, pp. 275--294, 1987.

\bibitem{estabrooks2000physical}
P.~A. Estabrooks and A.~V. Carron, ``The physical activity group environment
  questionnaire: An instrument for the assessment of cohesion in exercise
  classes.'' \emph{Group Dynamics: Theory, Research, and Practice}, vol.~4,
  no.~3, pp. 230--243, 2000.

\bibitem{chang2014high}
S.~Chang, L.~Jia, R.~Takeuchi, and Y.~Cai, ``Do high-commitment work systems
  affect creativity? {A} multilevel combinational approach to employee
  creativity.'' \emph{Journal of Applied Psychology}, vol.~99, no.~4, p. 665,
  2014.

\bibitem{chin1999perceived}
W.~W. Chin, W.~D. Salisbury, A.~W. Pearson, and M.~J. Stollak, ``Perceived
  cohesion in small groups: Adapting and testing the perceived cohesion scale
  in a small-group setting,'' \emph{Small group research}, vol.~30, no.~6, pp.
  751--766, 1999.

\bibitem{warkentin1997virtual}
M.~E. Warkentin, L.~Sayeed, and R.~Hightower, ``Virtual teams versus
  face-to-face teams: an exploratory study of a web-based conference system,''
  \emph{Decision Sciences}, vol.~28, no.~4, pp. 975--996, 1997.

\bibitem{dawes2008data}
J.~Dawes, ``Do data characteristics change according to the number of scale
  points used? {A}n experiment using 5-point, 7-point and 10-point scales,''
  \emph{International journal of market research}, vol.~50, no.~1, pp. 61--104,
  2008.

\bibitem{koo2016}
T.~K. Koo and M.~Y. Li, ``A guideline of selecting and reporting intraclass
  correlation coefficients for reliability research,'' \emph{Journal of
  chiropractic medicine}, vol.~15, no.~2, pp. 155--163, 2016.

\bibitem{rohlfing2019multimodal}
K.~J. Rohlfing, G.~Leonardi, I.~Nomikou, J.~Raczaszek-Leonardi, and
  E.~H{\"u}llermeier, ``Multimodal turn-taking: motivations, methodological
  challenges, and novel approaches,'' \emph{IEEE Transactions on Cognitive and
  Developmental Systems}, 2019.

\bibitem{dhall2015video}
A.~Dhall, O.~Ramana~Murthy, R.~Goecke, J.~Joshi, and T.~Gedeon, ``Video and
  image based emotion recognition challenges in the wild: Emotiw 2015,'' in
  \emph{Proceedings of the 2015 ACM on International Conference on Multimodal
  Interaction}, ser. ICMI '15.\hskip 1em plus 0.5em minus 0.4em\relax New York,
  NY, USA: ACM, 2015, pp. 423--426.

\bibitem{carron2000cohesion}
A.~V. Carron and L.~R. Brawley, ``Cohesion: Conceptual and measurement
  issues,'' \emph{Small group research}, vol.~31, no.~1, pp. 89--106, 2000.

\end{thebibliography}
\end{document}